# A DUAL-FUSION SEMANTIC SEGMENTATION FRAMEWORK WITH GAN FOR SAR IMAGES

*Donghui Li[1], Jia Liu[1*], Fang Liu[1], Wenhua Zhang[1], Andi Zhang[1], Wenfei Gao[1], Jiao Shi[2]*

1. Jiangsu Key Laboratory of Spectral Imaging & Intelligent Sense, Nanjing University of Science and Technology, Nanjing, 210094, China
2. Northwestern Polytechnical University, Xian, 710072, China

## ABSTRACT

Deep learning based semantic segmentation is one of the popular methods in remote sensing image segmentation. In this paper, a network based on the widely used encoder-decoder architecture is proposed to accomplish the synthetic aperture radar (SAR) images segmentation. With the better representation capability of optical images, we propose to enrich SAR images with generated optical images via the generative adversative network (GAN) trained by numerous SAR and optical images. These optical images can be used as expansions of original SAR images, thus ensuring robust result of segmentation. Then the optical images generated by the GAN are stitched together with the corresponding real images. An attention module following the stitched data is used to strengthen the representation of the objects. Experiments indicate that our method is efficient compared to other commonly used methods.

## 1. INTRODUCTION

Image segmentation plays a significant role in the interpretation of remote sensing data. Its goal is to assign image category labels to each pixel in the image [1]. SAR has become one of the important means of earth observation [2]. Remote sensing optical images are grayscale maps which contain multiple bands, and different bands characterize different targets, so they are advantageous for identification, classification, and extraction of target. Although SAR images are not influenced by the weather conditions, they are deeply affected by noise, and coupled with the typical geometric deformation of SAR images, e.g., layover, fore- shortening, multipath false targets. Generating accurate segmentation results from SAR images is more difficult than that of optical ones. Hence, how to effectively combine SAR images and optical images remains a key issue [3].

In recent years, a lot of semantic segmentation methods based upon deep learning have been popularly used for image segmentation tasks. U-Net [4] was originally used to solve medical image segmentation tasks, and the skip connection and U-shaped structure of it make good use of high-level semantic information and low-level features. Kaiming He recognized the phenomenon of degradation and proposed ResNet [5] to largely eliminate the difficulty in training over-deep networks. Liang-Chieh Chen utilizes Atrous Spatial Pyramid Pooling (ASPP) to solve multiscale problems [6]. Yu uses the Otsu method and level set to achieve segmentation of SAR images [7]. For the processing of Gaofen-3 SAR images, the modified Deeplabv3+ is used by Shi [8]. Zhu [10] used adversarial neural networks to implement style migration.

Although semantic segmentation methods based on deep learning result in good visual performance in the SAR images, they cannot be well used in the segmentation of targets with regular shape. As discussed above, optical images represent more details of the ground objects. Hence, this paper proposes an efficient network to reduce the influence of noise in SAR images via generated optical images. We add a generator to the front end of the encoder to generate the optical images corresponding to original images. The generator is trained via numerous SAR and optical images which can well represent the relationship between the two types of data and complement more details from the training data. The input data of encoder is the stitching of these optical images and original SAR images of the same size. The encoder is used to extract features to provide extraordinary information reference for decoder network, and the decoder is used to generate good results of segmentation.

In addition, establishing the dependencies between image pixels and attaining rich context information is still essential in semantic segmentation. To better capture the point, we add an effective module [9] related to attention mechanism after the GAN to strengthen the weight on the target.

This work was supported in part by the National Nature Science Foundations of China (Grant Nos. 61906093, 61802190, and 62076204), in part by Open Research Fund in 2021 of Jiangsu Key Laboratory of Spectral Imaging & Intelligent Sense (Grant No. JSGP202101), in part by the Nature Science Foundations of Jiangsu Province, China (Grant No. BK20190451), in part by the Fundamental Research Funds for the Central Universities (Grant No. JSGP202204), and in part by the Postdoctoral Science Foundation of Shannxi Province (Grant No. 2017BSHEDZZ77).

*Corresponding author: Jia Liu. Email: omegaliuj@gmail.com

The remaining sections of this paper are organized as follows. The details of our framework are specified in Section 2. Section 3 starts with the dataset used for the experiments and the details of experiments, followed by the result of ablation experiments and the comparison with other methods. In Section 4, our conclusion and the future work are stated.

## 2. METHODOLOGY

In this section, we first state the general architecture of our network. Afterwards, the generator used in the network, the attention module, and the encoder-decoder network are presented respectively.

### 2.1. Network Architecture

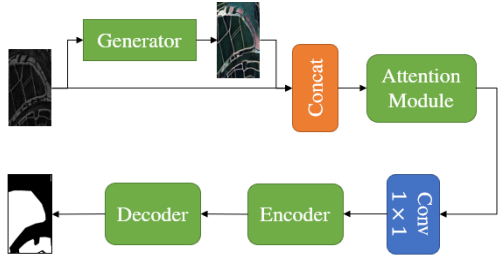

Fig 1: Network overall architecture.

The overall architecture of our network is illustrated in **Fig** 1. Before the original SAR images are input to the main part of the network, there is a branch for processing the images to generate approximate optical images corresponding to them. Subsequently, the result of this branch is merged with the original images in terms of number of channels. After this, the attention module is used to establish the dependencies between image pixels. The pointwise convolution is used to reduce the dimension of feature maps to match the input requirements. The data is then fed into the encoder-decoder network to attain the segmentation result. In those models, the optical image generator plays the critical role which is trained to render the SAR image with more details.

### 2.2. The Optical Images Generator

Remote sensing images are mostly acquired by media such as satellites, and therefore are difficult to produce and obtain paired data. Considering the situation, the GAN is based on the cycle-consistent adversarial network (CycGAN) [10], in order to enable the generation of images corresponding to the original input images in the case of using two different styles of image sets.

The structure of GAN is as follows in **Fig.** 2. In this figure, $X$, $Y$ are two sets of different styles of image collections. The network has two sets of generators $G_X$, $G_Y$ and discriminators $D_X$, $D_Y$. The generators generate images in the same style as the real images and the discriminators compare the results with the real images and score the results. The generators and discriminators are trained separately. When the parameters of the generators are fixed to train the discriminators, they can improve their ability. When the generators are trained with fixed parameters, they produce higher quality images to trick discriminators. Both sides evolve in an iterative learning process, eventually achieving dynamic balance where the generator produces an image close to the real one and the discriminator scores the image close to 0.5, because it cannot identify the image as true or false.

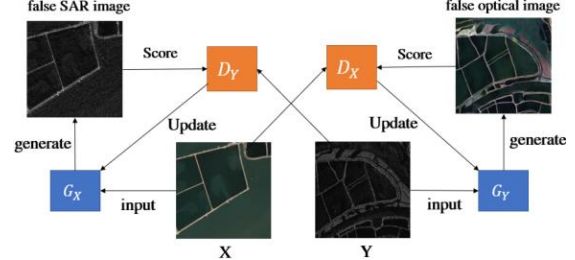

Fig. 2: GAN structure.

After the GAN has been trained, we need to introduce the generator model and associated parameters to transform the original input images when the framework is run.

### 2.3. The Attention Module

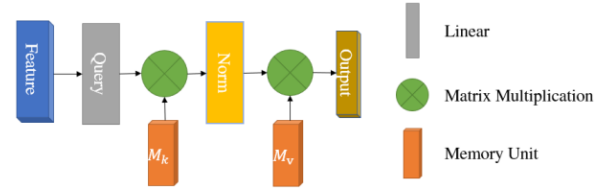

Fig 3: The Sketch of Attention Module.

For neural networks, attention mechanisms can focus on information that is more critical to the current task in the input information, reduce the attention to other information, and even filter useless information. The attention module of us is inspired by external attention [9]. A sketch of the attention module used in our network is shown in **Fig** 3. We first obtain the attention map by computing the pairwise associations between the self-query vector and the key-memory, and then the refined feature map is derived by counting the weights on this attention map from the value-memory. The output of the attention module is shown in the equation:

$$F_{out} = Norm\left(FM_k^T\right)M_v \qquad (1)$$

where $F$ is the input feature map, $M$ is learnable parameters that play the role of memory and are used to increase the network capacity, in which, $M_k$, $M_v$ are the key-memory and value-memory respectively, and $Norm$ in this figure refers to the operation of normalizing rows and columns separately. The memory units are accomplished by linear layers. They are independent of individual specimens and common throughout the dataset, which acts as a robust regularizer and improves the ability of the attention module to generalize.

## 2.4. Encoder-Decoder Network

The encoder-decoder is a common network structure whose idea of solving the image segmentation problem lies in the use of encoders for feature extraction and decoders for detail restoration.

Model Scaling has always been an important method to improve the effectiveness of convolutional neural networks. The size of the model is mainly determined by parameters in three aspects: width, depth, and resolution. For the high-resolution input images, a deeper network is needed to obtain larger receptive fields and use more channels to obtain finer features. We use compound scaling to achieve optimization of efficiency and accuracy by considering performance and arithmetic resources. We apply a form similar to EfficientNet-B7 [11] to implement the encoder stage for image feature extraction.

In the model, the decoder module is only based on Deeplabv3+ [12], because the body of the model is one of deep convolutional neural networks (DCNN) with atrous convolutions, which can control the receptive field without changing the feature map sizes, and is conducive to extract multiscale features. In addition, the atrous spatial pyramid pooling (ASPP) is more conducive to feature fusion and finally achieves an excellent segmentation effect.

## 3. EXPERIMENT

To assess the effectiveness of our method for image segmentation tasks in scenes such as marine farms, we use the FAIR1M [13] dataset in our experiments. In this section, we present this dataset and implementation details, and we analyze the experimental results, which include comparisons with other typical methods as well as ablation experiments.

### 3.1 Dataset

FAIR1M is a large-scale dataset for the task of target detection and recognition in remote sensing images. In order to meet the needs of practical applications, the images in FAIR1M dataset are gathered from different sensors and platforms with spatial resolutions ranging from 0.3m to 0.8m. Each image is in the size range from $1000 \times 1000$ to $10,000 \times 10,000$ pixels and contains objects showing various scales, orientations, and shapes.

In our experiments, we mainly use the data collected by HISEA-1 and GF-3 satellites. The scenes cover large mariculture farms commonly found in coastal areas. These images are a part of the dataset FAIR1M and each image contains two types of scenes: the mariculture farms and the backgrounds. They are single-channel images, and the label images are also in this form, where the pixel value of farms is 255, the pixel value of backgrounds is 0, and the image size varies from 512 to 2048 pixels. The total number of data sets is more than 6000.

We divided the dataset into the training set containing 3200 images, the validation set of 600 images, and used the remaining images as the test set.

### 3.2 Implementation details

Our network contains a style migration module based on the generators of CycGAN, and pretraining parameters of this module are obtained from training the adversarial network for 500 iterations. The input data of training is a set of 300 SAR images containing marine farms and a set of 50 optical images covering the same scenes.

We use one RTX3060 for experiments on the environment of Pytorch-1.9 and CUDA 11.1, and set the training time as 100 epochs, and batch size of 8. Specifying the loss function as a composite function with the ratio of Dice loss and binary cross-entry loss with sigmoid function is 1:3. The former is used to calculate the similarity of two samples whose formula is

$$L = 1 - \frac{2 * |X \bigcap Y| + 1}{|X| + |Y| + 1} \qquad (2)$$

where $X$ and $Y$ represent respectively the predicted maps and the ground truth maps, and $|X \cap Y|$ corresponds to multiplying the scores of each category in $X$ with the target in $Y$ and adding the elements of the result. The numerator and denominator are both added by 1 to avoid the problem of dividing the numerator by 0 as well as to reduce over-fitting. The latter is one of the common loss functions. AdamW is used as the optimizer for experiments. The initial learning rate is 0.01, the value of weight decay is $5 \times 10^{-4}$, and the minimum learning rate is set to $1 \times 10^{-5}$.

### 3.3 Results

In order to verify the results of models, the frequency weighted intersection over Union (FwIoU) is used as the evaluation index of the experiment for more accurate representation of categories. It is a kind of enhancement of the intersection over Union (IoU), and the weight can be set for each class according to its frequency.

Table 1: Results

| Encoder | Decoder | CycGAN | Attention | Combine | FwIoU |
|---|---|---|---|---|---|
| ResNet-34 | FPN | | | | 99.062 |
| | Deeplab v3 | | | | 98.828 |
| | Deeplab v3+ | | | | 98.986 |
| | Deeplab v3+ | √ | | | 99.063 |
| EfficientNet-b7 | Deeplab v3+ | | | | 98.761 |
| | Deeplab v3+ | √ | | | 98.786 |
| | Deeplab v3+ | √ | √ | | 99.094 |
| | Deeplab v3+ | √ | √ | √ | 99.118 |

As can be seen in **Table 1**, we consider the comparison experiment and the ablation experiment together, the check mark indicates that the module or the policy is chosen. Our network ultimately achieves the FwIoU of 99.118, which is better than others. The comparison of the effect of ablation experiments is shown in **Fig 4**. In this figure, 'Body' indicates the image segmentation using the encoder-decoder architecture only, 'Opt' indicates the process of the original image using the optical generator, and 'Att' indicates the addition of the attention module. According to the comparison of the experimental index and the comparison of the effect, our method is feasible.

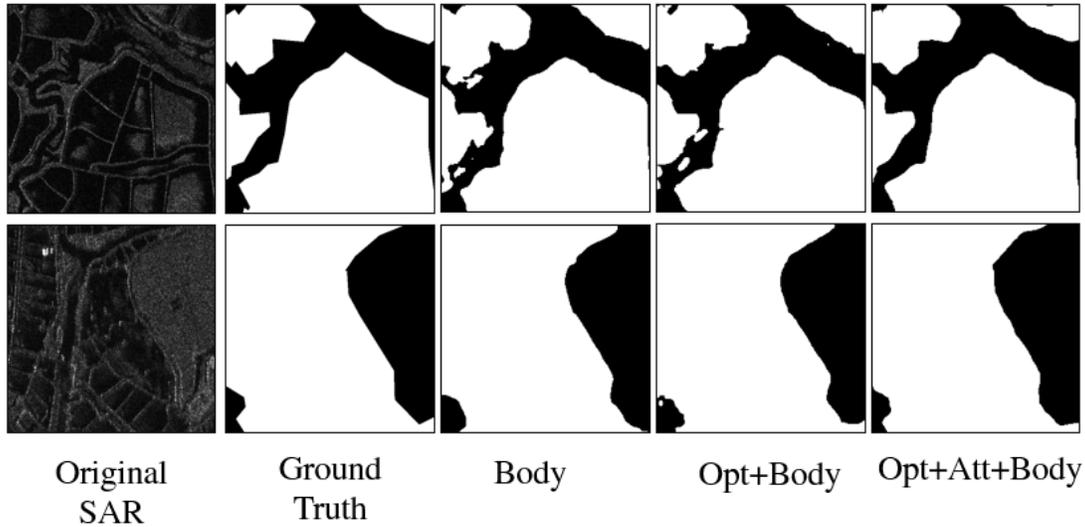

Fig. 4: The result of Ablation Experiments.

## 4. CONCLUSION

In this paper, we propose an image segmentation network using encoder-decoder architecture. The network uses style transformation to enhance the semantic expression of the input images, i.e., to transform the original SAR images into optical images. And an attention module is introduced to strengthen the contextual connection. The experiments have demonstrated that our approach performs well in the task of segmenting regular shaped objects such as marine farms. In the future work, we will integrate the generator into the network and try to enhance it in terms of clarity and recognizability of the generated images, updating it interactively with the network.